# Advanced Deep Learning Techniques for Accurate Lung Cancer Detection and Classification

Mobarak Abumohsen
*University of Vigo*
*Information Technologies Group,*
*School of Telecommunication*
*Engineering*, 36310 Vigo, Spain
mobarak.abumohsen@uvigo.gal

Enrique Costa-Montenegro
*University of Vigo*
*Information Technologies Group,*
*School of Telecommunication*
*Engineering*, 36310 Vigo, Spain
kike@gti.uvigo.es

Silvia García-Méndez
*University of Vigo*
*Information Technologies Group,*
*School of Telecommunication*
*Engineering*, 36310 Vigo, Spain
sgarcia@gti.uvigo.es

Amani Yousef Owda*
*Arab American University*
*Faculty of Graduate Studies*
*Department of Natural, Engineering*
*and Technology Sciences*
Ramallah, West Bank, Palestine
amani.owda@aaup.edu

Majdi Owda
*Arab American University*
*Faculty of Atrifical Intelligence*
*and Data Science*
*UNESCO Chair in Data Science for*
*Sustainable Development*
Ramallah, West Bank, Palestine
majdi.owda@aaup.edu

*Abstract*— Lung cancer (LC) ranks among the most frequently diagnosed cancers and is one of the most common causes of death for men and women worldwide. Computed Tomography (CT) images are the most preferred diagnosis method because of their low cost and their faster processing times. Many researchers have proposed various ways of identifying lung cancer using CT images. However, such techniques suffer from significant false positives, leading to low accuracy. The fundamental reason results from employing a small and imbalanced dataset. This paper introduces an innovative approach for LC detection and classification from CT images based on the DenseNet201 model. Our approach comprises several advanced methods such as Focal Loss, data augmentation, and regularization to overcome the imbalanced data issue and overfitting challenge. The findings show the appropriateness of the proposal, attaining a promising performance of 98.95% accuracy.

1. INTRODUCTION

Cancer poses a critical public health issue that is on the rise globally. This disease causes specific cell tissues to proliferate uncontrollably, ultimately this leads to malignant or tumor growth [1]. In 2020, GLOBOCAN recorded an anticipated 19.3 million new cancer instances with around 10 million deaths worldwide [2]. Additionally, the American Cancer Society projected that there would be 1,898,160 new cancer cases in the United States, leading to approximately 608,570 fatalities [3]. In Africa, there were 1,109,209 new cancer diagnoses and roughly 711,429 deaths that same year, with cancer death rates surpassing the combined fatalities from AIDS, tuberculosis, and malaria [4], [5].

Specifically, Lung Cancer (LC) ranks among the most frequently diagnosed cancers and is the major cause of death for both men and women worldwide. Annually, around 2.2 million new cases of LC are identified, resulting in approximately 1.8 million deaths globally [2] [6]. LC typically has a variety of typical signs and symptoms, including loss of appetite, fatigue, weight loss, chest pain, cough, shortness of breath (i.e., dyspnea), and hemoptysis [7]. Several risk factors are also linked with LC, including secondhand smoke, smoking, air pollution, radon exposure, and chemical exposure [7]. LC is divided into two broad categories: Small Cell LC (SCLC) and Non-Small Cell LC (NSCLC). The latter accounts for more than 80 percent of all LC and is then subdivided into three: Adenocarcinoma (ADC), Squamous Cell Carcinoma (SCC), and Large Cell Carcinoma (LCC) [8].

The imaging methods most commonly used in the detection of LC are Computed Tomography (CT), Positron Emission Tomography/Computed Tomography (PET/CT), and Magnetic Resonance Imaging (MRI) [9]. CT scans are often used in emergency settings because they take less time to process and cost less. Deep learning (DL), a recently emerging but fast-evolving branch of artificial intelligence, has the potential to revolutionize LC diagnosis and treatment. DL algorithms can be taught to recognize LC with high accuracy even in complicated cases using large sets of medical images.

Several studies [10], [11] have employed Machine Learning (ML) and DL techniques extensively for LC detection. Despite the promising results so far, accurate detection of LC is still a serious challenge [12]. The first is the class imbalance data, which negatively affects the performance and outcomes of the models. The second is model overfitting, which produces false results. In this paper, our aim is to overcome these challenges and make a contribution towards the development of a robust LC detection system from CT images to help medical professionals and radiologists make informed decisions on lung treatment and prevention by enabling reliable early detection. The primary contributions of this work are as follows:

- Build an efficient DL model to classify LC from CT images with high accuracy.
- Implement mechanisms such as data augmentation and Focal Loss to handle the imbalanced input data and

- improve classification performance across all LC classes.
- Utilize dropout layers, early stopping, and learning rate scheduling to minimize overfitting and ensure robust performance.

The rest of the article is structured as follows. Section 2 discusses previous research, whereas Section 3 describes the approach followed to create ML models for LC classification. Section 4 discusses the experimental findings and their comparison to previous works. Ultimately, Section 5 summarizes the findings of the study and provides recommendations for further research.

## 2. Literature Review

The ML and DL algorithms have attracted the interest of researchers for developing LC classification and detection models. Notably, ML, specifically DL, has emerged as a significant tool in medical imaging, transforming how healthcare practitioners approach illness detection and diagnosis [13], [14]. DL approaches, particularly convolutional neural networks (CNNs), have exhibited excellent performance in medical imaging applications [15]. Notably, VGG16, VGG19, DenseNet201, and InceptionV3 were selected for this study because of their shown efficacy in identifying and classifying multiple diseases from medical images, as evidenced by multiple previous studies [10], [11], [16]. CNNs are well-suited for evaluating CT, PET, and X-ray images because they learn spatial hierarchies of information, from edges in early layers to complex forms in in-depth layers, allowing them to reliably detect tumors and recognize between malignant and non-cancerous cells [17], [18].

Many researchers used a public or private CT image dataset to classify or detect LC using ML models. Authors in [11] proposed a VER-Net model for the LC classification from CT images. The proposed VER-Net model stacks three different transfer-learning algorithms, namely, EfficientNetB0, VGG19, and ResNet101. The proposed model attained over 90 % accuracy, precision, recall, and F1-score. Also, authors in [19] proposed AtCNN-DenseNet-201 TL-NBOA-CT for LC classification. The suggested model utilizes Modified Sage-Husa Kalman Filtering (MSHKF) for pre-processing, improved empirical wavelet transform (IEWT) for feature extraction, and attention-based CNN with DenseNet-201 for classification. Their proposed approach recorded an accuracy of 99.43 %. Furthermore, researchers in [20] developed a novel DL algorithm to classify NSCLC based on a CT imaging dataset. Dense neural networks (VGG-16 and Resnet-50) and sparse neural networks (Inceptionv3) were applied on pictures of 60 ADC and SCC patients, respectively, from a public dataset. Deep learning was employed for feature extraction from the CT images. The results show that the Inceptiov3 model achieved the best accuracy of 98.29 %. Authors in [21] designed a new ML algorithm to identify stage I-IIIA NSCLC from image data acquired through CT scans. The hybrid model proposed is a picture and clinical information-based hybrid model, which was constructed using DL 3D CNN. The median AUC of the hybrid model proposed was 0.76.

In addition, there existed certain studies that used a public CT dataset from Kaggle. The authors of [16] developed a new ML approach to NSCLC detection using a public CT imaging database (LIDC-IDRI). The authors used traditional ML for LC detection (Support Vector Machine (SVM), Random Forest (RF), K-Nearest Neighbors (KNN), Decision Tree (DT)), and additional methods for LC classification such as the VGG19, EffetientNet-V2-L, Wide-Resnet-50-2-weight, and EffetientNet-B7 models. The highest accuracy was achieved by the VGG19 model at 99.7 %, followed by EfficientNet-V2-L with 99.31 %, and traditional ML had low accuracy. Besides, the authors in [22] presented a DL model for the classification of four classes of LC using publicly available CT scan images on Kaggel. The model achieved 96 % accuracy when it used the DL method EfficientNet B3 to identify CT scans as normal, SCC, and ADC. For instance, authors in [23] have used a new DL model for the identification of binary classification LC on also public CT image database on Kaggle. They are classifying with the model of CNN. A user-end app allows users to upload their CT images for them to be processed while the training data is employing lung CT data. The results indicate that the proposed CNN algorithm achieves an accuracy rate of 97.10 %.

Some studies utilized a public or private CT image dataset to classify and segment LC based on ML and DL technique. In [24], a new ML model was proposed for detecting and segmenting lung malignant tumors using CT images gathered at the Imperial College Hamlyn Center, London. The CNN model and some techniques were applied to, first, remove the noise and enhance the input images used. They then used edge-based segmentation and region of interest (ROI) algorithms to segment the image. The accuracy of this method was 99.80 %. In addition, researchers in [25] proposed a new DL model for LC classification and segmentation using an open CT dataset (LIDC-IDRI) scanned at the Foundation for the National Institutes of Health (FNIH). The suggested techniques included preprocessing, segmentation of the lung area, data augmentation, segmentation of the cancerous tumor, and classification of the lung for cancer detection. The Shepard Convolutional Neural Network (ShCNN) trained with the Water Cycle Sea Lion Optimization (WSLnO) algorithm was effectively used for LC classification. This model achieved an accuracy of 90.91 %.

The same dataset was utilized by researchers in [10] for our task to build a new DL model to diagnose NSCLC from a public dataset Kaggle based on CT image chests. They proposed a new method, a dual-state transfer learning (DSTL) method based on a deep CNN-based approach, to develop an efficient model that can classify the type of LC accurately, reduce variance, and avoid overfitting by leveraging the DCNN, VGG16, Inceptionv3, and RestNet50 models. This proposal attained an accuracy of 92.57 %.

According to the reviewed publications, there is a significant gap in research using Focal Loss to solve the class imbalance in LC classification. Furthermore, few studies have used approaches like dropout layers, early halting, and learning rate scheduling to effectively reduce overfitting while maintaining high model performance. This emphasizes the originality and relevance of incorporating these sophisticated approaches into the suggested methodology for LC diagnosis.

## 3. METHODOLOGY

The process followed in this study is illustrated in Figure 1. It begins with data gathering, followed by preprocessing to prepare the images. Data augmentation and feature extraction using pre-trained models are then performed. Classification of lung cancer is conducted using VGG16, VGG19, InceptionV3, and DenseNet201. A comparison of the models is then made to select the best model.

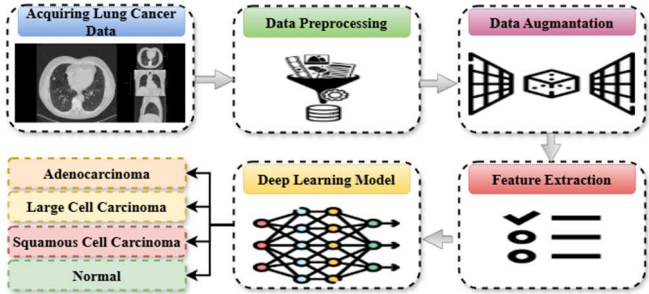

Figure 1: Workflow used for LC classification.

### 3.1 Data Selection

This study employed the publicly Chest CT Scan Images Dataset composed of lung CT images from Kaggle [26]. This dataset was primarily designed for medical imaging applications like LC classification. It contains 1,185 CT scan pictures divided into four categories: ADC (534), LCC (166), normal (240), and SCC (245).

### 3.2 Data Preprocessing

Data preparation seeks to standardize and improve the dataset quality in order to train DL models more efficiently [27]. The collection contains images of various resolutions, including 224×224, 199×199, 112×112, and others. To maintain consistency and compatibility with DL models, all scans were addressed to a common size of 299×299 pixels during preprocessing. The CT scan dataset was preprocessed by converting RGB images to grayscale, resizing them to a consistent 299×299 pixels using bilinear interpolation for compatibility with pre-trained models, and normalizing pixel values to the range [0, 1] to stabilize training and improve numerical stability. Class labels were encoded numerically (ADC: 0, LCC: 1, Normal: 2, SCC: 3) for classification purposes. The preparation procedure, as seen in Figure 2, resulted in two outputs: a numpy array of standardized grayscale images and another array of numerical labels. The advanced techniques and libraries were utilized to ensure effective preprocessing and data preparation, including OpenCV (cv2.IMREAD) and OpenCV (cv2.resize), Pixel normalization by dividing values by 255.0, and TensorFlow (to_categorical).

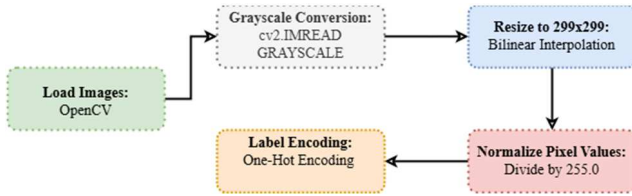

Figure 2: Workflow for data preprocessing.

### 3.3 Data Augmentation

Data augmentation is used to increase the variety of the training dataset while also improving the DL model's generalization capabilities. This stage is particularly important in medical imaging, where datasets are frequently tiny and unbalanced. By varying the current data, the model becomes more resilient and less prone to overfitting. Figure 3 shows the technique used for the data augmentation.

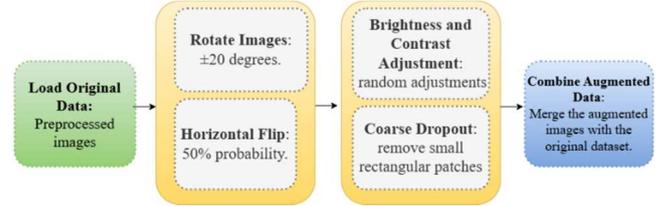

Figure 3: Data augmentation technique.

Figure 3 shows the data augmentation step, which involves rotation, flipping, brightness modification, and coarse dropout. These methods increase data variety, model generalization, and decrease overfitting in LC classification.

### 3.4 Feature Extraction

Feature extraction uses pre-trained DL models to extract significant high-level characteristics from CT scan pictures [28], [29]. This stage takes advantage of transfer learning, in which models trained on big datasets such as ImageNet are applied to the LC classification problem. Using pre-trained models eliminates the requirement for labeled medical data while maintaining high accuracy. Figure 4 shows the workflow used for the feature extraction.

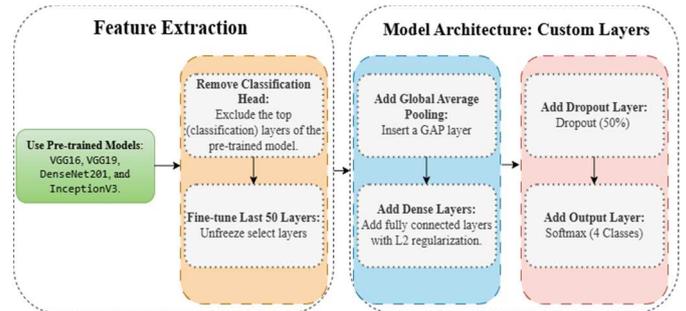

Figure 4: Workflow used for the feature extraction.

Figure 4 shows the procedure for feature extraction. Pre-trained models are fine-tuned by deleting the classification heads and unfreezing certain layers. Custom layers, such as global average pooling, dense layers, dropout, and a Softmax output layer, are used to classify LC into four groups, enclosing high accuracy.

### 3.5 Data Model Training

The model is trained to utilize features from pre-trained networks and custom layers, with sophisticated optimization approaches used to solve class imbalance and overfitting. Carefully chosen tactics provide robust learning and efficient convergence, as described in the training procedures that follow.

#### 3.5.1 Loss Function

Focal loss is used to resolve class imbalance by directing more attention to minority classes data [30]. This adaptive loss function favors difficult cases during training, ensuring that underrepresented classes make important contributions to model optimization.

### 3.5.2 Optimization (Adam Optimizer, Learning Rate Scheduler)

The Adam optimizer is used given its efficiency in gradient-based optimization. It adjusts the learning rate dynamically during training, resulting in quicker and more consistent convergence than standard optimizers. A dynamic learning rate scheduler adjusts the learning rate as a function of validation performance [31]. During plateauing in validation loss, the learning rate is reduced in order to allow the model to experiment with diminishing changes, leading to enhanced performance overall.

### 3.5.3 Callbacks (Early Stopping, Reduce Learning Rate on Plateau)

Early stopping is used to avoid overfitting by monitoring validation loss and halting training after a given number of epochs with no improvement[32]. This guarantees that the model does not lose its generalizability. When the validation performance plateaus, the learning rate automatically decreases, allowing the model to fine-tune learning and avoid potential local minima. The training process, spurred by features learned from pre-trained networks and aided by aggressive optimization techniques combination of Adam optimizer, focal loss, learning rate scheduling, and early stopping, offers great accuracy and stability in LC CT image classification. The use of Focal Loss and dynamic callbacks tailors the training process to the specific challenges of medical image analysis.

### 3.6 DL algorithms

This section describes the DL models used for feature extraction and classification of LC using a CT scan dataset. VGG16, VGG19, DenseNet201, and InceptionV3 models are employed for feature extraction and classification.

### 3.6.1 VGG16

The VGG16 model is made up of 16 layers and gained massive popularity by claiming the ILSVRC-2014 challenge. It has over 138 million parameters and is built with standard convolutional layers, same padding, and max-pooling layers. Convolutional layers employ $3 \times 3$ filters with a one-stride, while max-pooling layers employ $2 \times 2$ filters with a two-stride [33]. The model has 13 convolutional and max-pooling layers, well stacked along. The input layer receives 299×299 pixel images. The network ends with fully connected layers, with two Rectified Linear Unit (ReLU) activation functions and the final SoftMax activation for classification.

### 3.6.2 VGG19

The VGG19 model, a variant on the VGG architecture, has 19 layers: 16 convolutional layers, 3 fully connected layers, 5 max-pooling layers, and 1 SoftMax layer. The processing of 299×299 RGB pictures requires 19.6 billion FLOPs. Pre-processing subtracts the average RGB value from each pixel. Convolutional layers employ 3×3 kernels with a stride of 1 and spatial padding to retain resolution. Max-pooling uses 2×2 windows and a stride of 2 for down sampling. ReLU activation increases accuracy while decreasing calculation time [34]. The design concludes with three fully connected layers: two with 4096 nodes and one with 1000 channels for ImageNet classification, followed by a SoftMax function for output probabilities to improve accuracy while remaining efficient.

### 3.6.3 DenseNet201

DenseNet201 is a well-known variant of the DenseNet architecture, consisting of 201 layers. The model's structure, converted into the h5 format, includes dense block 1, transition layer 1, dense block 2, transition layer 3, transition layer 4, and the classification layer [35]. It processes input images of size 299×299, enabling it to learn a wide variety of feature representations suitable for different image types.

### 3.6.4 InceptionV3

The InceptionV3 architecture, created by Google [36], improves on previous versions with features like as factorized convolutions, which minimize the amount of network parameters and hasten training. An auxiliary classifier is introduced to improve performance by functioning as a regularize, and pooling operations are conducted during grid size decreases. InceptionV3 includes symmetric and asymmetric convolution blocks, average and maximum pooling operations, concatenations, dropouts, and fully linked layers. The network has 42 layers and takes input images of 299×299 pixels. Near the conclusion, fully linked layers are introduced, followed by the final classification layer, SoftMax.

### 3.7 Model Evaluation

Following Once trained, the model performance is evaluated against the test dataset to determine its effectiveness and reliability. The evaluation method is optimizing major parameters and is comparing different backbones to determine the top-performing model for LC CT scan classification. The performance of every DL model ranging from Equation (1) to Equation (4) is evaluated based on the following metrics:

$$Accuracy = \frac{TP + TN}{TP + TN + FP + FN} \quad (1)$$

$$Precision = \frac{TP}{TP + FP} \quad (2)$$

$$Recall = \frac{TP}{TP + FN} \quad (3)$$

$$F1 - Score = 2 * \frac{Precision * Recall}{Precision + Recall} \quad (4)$$

Where TP: true positive, TN: true negative, FP: false positive, and FN: false negative.

## 4. RESULT AND DISCUSSION

This section discusses the result of the proposed methodology, which was experimented on according to different DL models (DenseNet201, InceptionV3, VGG16, and VGG19) for classifying LC. Performance is compared in terms of accuracy, recall, precision, F1 score, and training model responsiveness.

DenseNet201 outperformed the others with a satisfactory accuracy of 98.95 %, proving its superiority in feature extraction for LC classification. InceptionV3 was also well-performing, with an accuracy of 95.57 %, and is therefore a good choice for this task, although less accurate than DenseNet201. VGG16 and VGG19, however, achieved lower accuracies of 63.08 % and 23.42 %, respectively, which prove their failure to manage complex LC patterns and imbalanced datasets.

The training process shows the consistency of DenseNet201, as it continued to improve over epochs and eventually led to the highest accuracy among competing models. This demonstrates great generalization with little overfitting. Despite high variability during training, InceptionV3 delivered high validation accuracy and ranked second. VGG16 and VGG19, however, suffered from overfitting and poor generalization, represented by low validation accuracies and overall classification performance.

Table 1 shows DenseNet201 model to have had impressive performance across all classes with almost perfect precision, recall, and F1-scores. For ADC, the model scored 100% precision and 99% recall and thus 100% F1-score. Similarly, equally impressive values were found in LCC, Normal, and SCC classes where 98%, 98%, and 99% F1-scores were reported. The results identify DenseNet201's potential for accurate LC subtype classification and its applicability in clinical practice.

TABLE 1: DenseNet201 Classification Metrics.

| Class | Precision % | Recall % | F1-Score % |
|---|---|---|---|
| ADC | 100 | 99 | 100 |
| LCC | 97 | 98 | 98 |
| Normal | 98 | 99 | 98 |
| SCC | 99 | 99 | 99 |
| Overall | 99 | 99 | 99 |

Figure 5 shows the training and validation accuracy of the DenseNet201 model over 20 epochs, exhibiting its good learning and generalization ability.

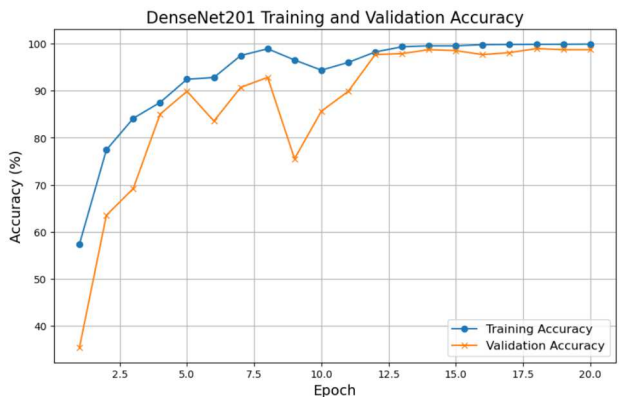

Figure 5: DenseNet201 training and validation accuracy.

Figure 6 illustrates the performance comparison of precision, accuracy, F1-score, and recall of DenseNet201, InceptionV3, VGG16, and VGG19 models. DenseNet201 performed highest across all measures followed by InceptionV3, with VGG16 and VGG19 performing significantly lower, an indication of DenseNet201 and InceptionV3's improved feature extraction and classification ability for LC detection.

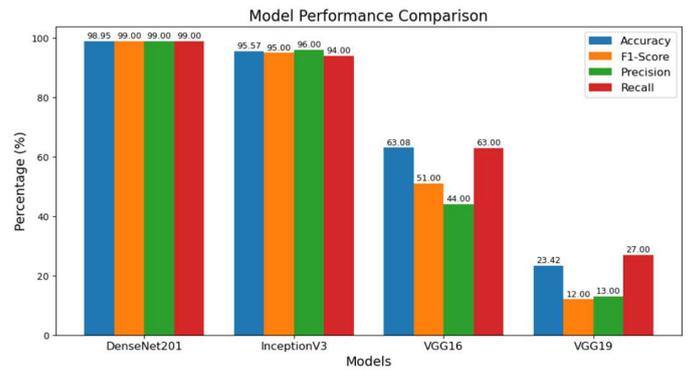

Figure 6: Performance comparison of each classification model.

Table 2 shows a comparison of performance with existing studies, and the suggested model achieves the best accuracy of 98.95 %. This is due to sophisticated methods like focal loss, learning rate scheduling, and early stopping, which improve model robustness and generalization.

TABLE 2: Comparison with Previous Research.

| Ref | Dataset | Method | Accuracy |
|---|---|---|---|
| [11] | Chest CT Scan | VGG19 + EfficientNetB0 + ResNet101 | 91% |
| [22] | | EfficientNet B3 | 96% |
| [10] | | Dual-state transfer learning (DSTL), DCNN, VGG16, Inceptionv3, and RestNet50. | 92.57% |
| Our study | | Proposed approach | 98.95% |

## 5. Conclusions

This study introduced a new DL-based model for LC classification from CT scans. The method introduced addressed the pertinent problems: model overfitting and data imbalance using various unconventional techniques such as regularization, focal loss, and data augmentation. The outcomes demonstrated that DenseNet201 outperformed its counterparts (InceptionV3, VGG16, and VGG19); it attained an accuracy of 98.95%, precision of 99%, recall of 99%, and F1-score of 99%. Moreover, this study conducted a comparison of the performance of the proposed approach with the cutting-edge techniques employed in previous studies and showed impressive efficacy compared with them.

Our future directions include evaluating the proposed approach on different datasets toward a generalized model, modifying the suggested approach to include the segmentation process, and finally investigating a new approach for the feature extraction and selection process.